\documentclass[structabstract]{aa}  
\usepackage{graphicx}
\usepackage{txfonts}
\begin{document}

   \title{The role of black hole spin and magnetic field threading the unstable neutrino disk in Gamma Ray Bursts}

   \author{Agnieszka Janiuk\inst{1}
          \and
Ye-Fei Yuan\inst{2}
}


   \institute{Copernicus Astronomical Center, Polish Academy of Sciences, ul. Bartycka 18, 00-716 Warsaw, Poland \\
\email{agnes@camk.edu.pl}
\and
 Key Laboratory for Research in Galaxies and Cosmology, University of Science and Technology of China, Chinese Academy of Sciences, Hefei 230026,
China\\
\email{yfyuan@ustc.edu.cn}
}
   \date{Received ...; accepted ...}
  \abstract
{}
{We report on the third phase of our study of the neutrino-cooled hyperaccreting
torus around a black hole that powers the jet in Gamma Ray Bursts.
We focus on the influence of the black hole spin on the properties of the torus.}
   {The structure of a stationary torus around the Kerr black hole 
is solved numerically. We take into account the detailed treatment of the microphysics in the nuclear equation of state that includes the neutrino trapping effect. 
}
{We find, that in the case of rapidly rotating black holes, the thermal instability
discussed in our previous work is enhanced and develops for much lower accretion rates.
We also find the important role of the energy transfer from the rotating black hole to 
the torus, via the magnetic coupling.
}
{}

  \authorrunning{Janiuk \& Yuan}
  \titlerunning{Unstable GRB disk around a spinning  black hole}
   \maketitle

\section{Introduction}

Observed Gamma Ray bursts (GRBs) form at least two classes of events. 
Long GRBs ($T_{90} > $ 2 s) are the the result of a collapsar in the core of an 
exploding massive star, as supported by the detections of their afterglows. 
Shorter bursts ($T_{90} < $ 2 s) are likely to be due to the coalescence of a binary 
system containing two compact stars, as observationally supported by the 
localization of their afterglows at the outskirts of their host galaxies (see e.g. 
the review of Zhang 2007). 
Moreover, it is possible that a third class of events exists, namely a fraction of 
the shortest bursts ($T_{90} < $ 0.1 s), that can be produced 
via the evaporation of the
primordial black holes (Cline et al. 2005).

Both the two main classes of bursts involve a stage of a newly formed black 
hole surrounded by an accreting torus, responsible for the production of
a collimated jet. 
This jet is a source of a highly beamed emission observed in the gamma ray band.
The accreting torus is either fed by the material 
from the stellar envelope, which part falls back after the hypernova explosion, or 
consumes the material from the disrupted neutron star debris.
The torus accretion proceeds with highly hyper-Eddington rates, on the order of
a solar mass per second.
Such hyper-accreting tori have been already discussed in a number of works
(e.g. Popham et al. 1999; Di Matteo et al. 2002; Kohri et al. 2005; 
Chen \& Beloborodov 2007; Janiuk et al. 2004; 2007; Lei et al. 2009).

In the present paper we expand on our previous work and we further 
investigate the properties and evolution of such hot and dense 
accretion tori.
At the extreme densities and temperatures, determined by the hyper-Eddington
 accretion rate, the 
torus is cooled mainly by neutrino emission produced primarily by electron 
and positron capture on nucleons ($\beta$ reactions). 
The model we develop here was first described in Janiuk et al. (2004) for the case of a 
non-rotating black hole and the nuclear matter with a simplified 
equation of state.
We further developed this model in Janiuk et al. (2007), considering the more 
elaborate equation of state to account for the microphysics of the accreting plasma. 
In the latter work, 
we solved for the disk structure and its time evolution by introducing 
the new EOS which includes photodisintegration of helium, 
the condition of $\beta$ equilibrium, and neutrino opacities. We self-consistently 
calculated the chemical equilibrium in the gas consisting of helium, free protons, 
neutrons, and electron-positron pairs and compute the chemical potentials of the 
species, as well as the electron fraction throughout the disk. 
We found an important property of our solutions: for very large accretion rates, 
the torus becomes viscously 
and thermally unstable in a narrow range of radii near the central black hole. 
The instability results as the intrinsic property of the torus, when the helium is 
fully photodisintegrated in the neutrino opaque plasma.

In the present work, we further expand our model to account for the black hole 
rotation. The rotating black hole should naturally be produced in the center of a 
collapsar or after the merger event, when the collapsing material possesses a large 
amount of angular momentum. As a consequence, the black hole spin is a plausible 
mechanism that helps to launch the jets emitted in a narrow cone along the rotation 
axis. As shown in Janiuk et al. (2008), especially for the long duration GRBs the BH 
spin is required to sustain the central engine activity and the 
jet production for a sufficiently long time.
Because these long GRBs often exhibit very variable temporal structure (e.g. Beloborodov, Stern \& Svensson 2000), it 
seems very important to study a possible instability mechanism that may account for 
the variability.
We check here, for what black hole spin the instability may operate near the central 
black hole, depending on the accretion rate and viscosity in the disk. 
We also check what effects may be imposed on the torus, and its unstable strip, 
by the additional heating via the magnetic fields, threaded by the rotating black hole.
The content of the article is as follows.
First, we discuss the basic assumptions and present the main equations
of the model. Second, we introduce the changes and corrections to the model
that are the result of a black hole rotation with an arbitrary spin.
Third, we introduce the energy extraction from the rotating black hole
via the magnetic field, as an additional physical process that may operate
in the gamma ray burst central engine. 
In Section \ref{sec:results} we present the results of our calculations, and
in Section \ref{sec:diss} we discuss the results and conclude.

\section{Model}
\label{sec:model}

The model of a neutrino cooled accreting torus was fully 
discussed in Janiuk et al. (2004; 2007). 
Here we briefly repeat the basic assumptions and the model equations
are given in Sec. \ref{sec:basic}. In Section \ref{sec:kerr} we 
describe the relativistic corrections to the model equations, which account 
for the black hole rotation in the Kerr metric. 
In Section \ref{sec:qmc} we discuss the additional heating term in the torus energy 
balance, that comes from the rapid black hole rotation and magnetic dynamo.

\subsection{Basic equations}
\label{sec:basic}

The starting model for the torus evolution is calculated by solving the
energy balance equation at a given accretion rate, $\dot M$, black hole mass, $M$, and 
viscosity, $\alpha$.
The steady-state model is vertically averaged, with a surface density $\Sigma = H \rho$,
and height $H=c_{s}/\Omega_{K}$. Here the sound speed is $c_{s} = \sqrt{P/\rho}$, 
$P$ is the total pressure and $\Omega_{\rm K} = \sqrt{GM/r^{3}}$
is the Keplerian angular velocity. The viscosity is described as in Shakura \& Sunyaev 
(1973), with $\tau_{r\varphi}=-\alpha P$ and $0 < \alpha < 1$ is a model parameter.

Throughout the calculations we adopt a black hole mass of $M=3 M_{\odot}$ and 
several chosen values of $\alpha$ and $\dot M$.

For the equation of state, we  assume that the torus consists of helium, 
electron-positron pairs, free neutrons and  protons. The total pressure is 
due to the free nuclei and pairs, helium, radiation and the trapped 
neutrinos:
\begin{equation}
P = P_{\rm nucl}+P_{\rm He}+P_{\rm rad}+P_{\nu}\;.
\end{equation} 
where $P_{\rm nucl}$ includes free neutrons, protons, 
and the electron-positron pair gas in beta equilibrium, taking into account the 
partial degeneracy and relativistic temperatures of the species 
(Yuan 2005; Yuan \& Heyl 2005).
The helium is generally non-relativistic
and non-degenerate; therefore, its pressure is given by the ideal gas EOS.
Its number density is calculated using the formula of Popham et al. (1999)
\begin{equation}
X_{\rm nuc}=295.5\rho_{10}^{-3/4}T_{11}^{9/8}\exp(-0.8209/T_{11}).
\label{eq:xnuc}
\end{equation}
The radiation pressure includes the factor due to the electron-positron pairs, and
the neutrino pressure is calculated for the trapped neutrinos, using the 
two-stream approximation (Popham \& Narayan 1995; Di Matteo et al. 2002).
We consider both the neutrino transparent and opaque regions,
as well as the transition between the two, and in order to determine the distribution 
function of the partially trapped neutrinos we use a "gray body" model (Sawyer 2003), 
consistent with the two-stream approximation.

The chemical potentials, or equivalently  
the ratio of free protons, $x=n_{p}/n_{b}$, are determined from the 
condition of equilibrium between the transition reactions from 
neutrons to protons and from protons to neutrons
for a given baryon number density, $n_{\rm b}$, and temperature, $T$. 
These reactions are electron and positron capture on nucleons, and neutron decay 
(see Kohri, Narayan and Piran 2005; Janiuk et al. 2007).
The closing equations for the EOS are the conservation of the baryon number, 
$n_n+n_p=n_b X_{\rm nuc}$, and charge neutrality (Yuan 2005).

For the neutrino cooling of the torus, we consider the electron-positron pair 
annihilation, bremsstrahlung, plasmon decay and beta reactions. 
Each of these emission processes
 has a reverse one, which leads to neutrino absorption, and we calculate the 
absorptive optical depth $\tau_{\rm a,\nu_{i}}$ for the neutrinos of the three flavors. 
In addition, the free escape of neutrinos from the disc is limited by 
scattering, and we calculate the scattering optical depth $\tau_{\rm s}$.
The neutrino cooling rate is therefore in the neutrino-thick torus given by
\begin{equation}
Q^{-}_{\nu} = { {7 \over 8} \sigma T^{4} \over 
{3 \over 4}} \sum_{i=e,\mu} { 1 \over {\tau_{\rm a, \nu_{i}} + \tau_{\rm s} \over 2} 
+ {1 \over \sqrt 3} + 
{1 \over 3\tau_{\rm a, \nu_{i}}}}\;.
\label{eq:qnuthick}
\end{equation}

Apart from the neutrino emission, the disk is also cooled by advection,
radiation and photodissociation of helium nuclei.
The advective cooling is determined approximately as:
\begin{equation}
Q^{-}_{\rm adv}=  \Sigma v_{r} T {d S \over d r} = 
q_{\rm adv}{\alpha P H T \over \Omega \rho r^{2}} S\;,
\label{eq:fadv}
\end{equation}
 where $q_{\rm adv} \propto d\ln S/d \ln r \propto(d \ln T/d \ln r - (\Gamma_{3}-1)d \ln \rho/d \ln r)
  \approx 1.0$.
The entropy density $S$, similarly to the total pressure, is the sum of four 
components: 
\begin{equation}
S = S_{\rm nucl}+S_{\rm He}+S_{\rm rad}+S_{\nu}\;.
\end{equation}
and we calculate the entropy from the EOS.

The radiative cooling is 
\begin{equation}
Q^{-}_{\rm rad}={3 P_{\rm rad} c \over 4\tau}={11 \sigma T^{4} \over 4
\kappa \Sigma}
\label{eq:qrad}
\end{equation}
where we adopt the Rosseland-mean opacity
$\kappa=0.4+0.64\times 10^{23}\rho T^{-3}$ ~[cm$^{2}$g$^{-1}$], and the 
coefficient $11/4$ accounts for the contribution of electron-positron pairs.

The cooling rate due to the photodisintegration of $\alpha$ particles is:
\begin{equation}
Q^{-}_{\rm photo} =  6.28 \times 10^{28} \rho_{10} v_{r} {dX_{\rm nuc} \over dr} H
\label{eq:photodis}
\end{equation}
and $X_{\rm nuc}$ is given by Equation~\ref{eq:xnuc}.

Finally, the hydrodynamic equations we solve to calculate the disc structure are
the standard mass, energy and momentum conservation.
The viscous heating rate is determined by the shear, $\tau_{r \varphi}$
and can be written in terms of $\alpha$:
\begin{equation}
Q^{+}_{\rm visc}={3 \over 2}\alpha \Omega H P.
\end{equation}

The total flux of energy generated at the radius $r$ is determined by the global 
model parameters, i.e. the black hole mass and accretion rate:
\begin{equation} 
F_{\rm tot} = {3 G M \dot M \over 8 \pi r^3} f(r)
\label{eq:ftot}
\end{equation}
where $f(r)$ stands for the inner boundary condition and will be 
given below.
In order to calculate  the initial stationary  configuration, 
we solve the energy balance:
\begin{equation}
F_{\rm tot} = Q^{+}_{\rm visc} = Q^{-}_{\rm adv}+Q^{-}_{\rm
  rad}+Q^{-}_{\nu} + Q_{\rm photo}.
\end{equation}

The above equations are appropriate in the Schwarzschild black hole case, 
studied in our previous work (Janiuk et al. 2004; 2007), and must be
modified in the Kerr metric. In Sec. \ref{sec:kerr}, we include the 
appropriate correction factors to the disk structure equations to introduce the
Kerr metric and account for the black hole rotation.

\subsubsection{Instability in the disk}

The stability analysis of the neutrino-ccoled disk in $\beta$-equlibrium
was presented presented in Janiuk et al. (2007). The criterion for a viscously unstable 
disk is:
\begin{equation}
{\partial{\dot M} \over \partial{\Sigma}} < 0
\end{equation}
and means that, at a given radius, 
the underdense (overdense) region evolves slower (faster), which results
in material being evacuated from some rings of the disk, while being piled 
up in the others. The unstable region can be conveniently located 
in the steady-state solutions, on the 
surface density - accretion rate (or temperature) plane. On this plane, the 
branch of thermal equilibrium solutions, which has a negative slope, 
is unstable to the surface density perturbations. This branch is also thermally 
unstable, and we have
\begin{equation}
{\partial \log Q^{+} \over \partial \log T} > {\partial \log Q^{-} \over \partial \log T}
\end{equation}
and therefore
 any small increase (decrease) in temperature leads to more (less) heating rate 
with respect to the cooling.

We found, that for a non-rotating black hole, the neutrino-cooled disk was generally 
stable for all radii if $\dot M \le 10 M_{\odot}$s$^{-1}$, while at larger 
accretion rates the unstable region appears at the innermost radii.
The instability grows, because the energy balance changes because of the 
photodisintegration of helium, while the neutrinos are trapped in the opaque plasma.
In the unstable region, the electrons and protons become non-degenerate
and the electron fraction rises inwards in the disk. The changes in the electron 
fraction influence the total pressure and the energy energy dissipation rate, 
so that the heating and cooling balance is affected and the 
local accretion rate increases. We found that for the highest accretion rates we 
used (12 $M_{\odot}$s$^{-1}$) 
and the non-rotating black hole, the radial extension of the unstable
part of the disk
was up to $\sim 8$ Schwarzschild radii.

\subsection{Black hole rotation}
\label{sec:kerr}

The structure of the relativistic accretion disk is calculated 
using the correction factors derived in Riffert \& Herold (1995) for
a Kerr black hole (see e.g. Reynoso, Romero \& Sampayo 2006). 
The black hole spin is parameterized by
a dimensionless specific angular momentum (Bardeen 1970), 
$a=cJ/GM^{2}$, and the following 
functions of $a$, $M$, and radial coordinate $r$ are introduced:

\begin{equation}
A = 1 -{2 G M \over c^{2} r} + \left({GM a \over c^{2}r}\right)^{2}
\end{equation}
\begin{equation}
B = 1 - {3 G M \over c^{2} r } + {2 a \left({GM \over c^{2} r}\right)^{3/2}}
\end{equation}
\begin{equation}
C = 1 - 4 a \left({GM \over c^{2} r}\right)^{3/2} + 3 \left({GM a \over c^{2}r}\right)^{2}
\end{equation}
\begin{equation}
D = {1 \over 2 \sqrt{r} } \int^{r}_{r_{\rm ms}} {{x^{2}c^{4} \over G^{2}} - 6{M x c^{2} \over G} + 8a\sqrt{M^{3}x c^{2} \over G} - 3 a^{2}M^{2} \over \sqrt{x}\left({x^{2}c^{4} \over G^{2}} - 3 {M x c^{2} \over G} +2a\sqrt{M^{3}x c^{2} \over G}\right)} dx
\end{equation}

The above coefficients approach unity at large radii and for small spin parameters. The
inner boundary of the disk is located at the last stable circular orbit, depending 
on the black hole spin (Bardeen, Press \& Teukolsky 1972):
\begin{equation}
r_{\rm ms} = {GM \over c^{2}}(3+z_{2}\pm \sqrt{(3-z_{1})(3+z_{1}+2 z_{2})})
\label{eq:rms}
\end{equation}
where $z_{1}=1+(1-a^{2})^{1/3}((1+a)^{1/3}+(1-a)^{1/3})$, 
$z_{2} = (3 a^{2}+z_{1}^{2})^{1/2}$, and the upper sign in Eq.\ref{eq:rms} is for a 
retrograde orbit while the bottom sign is for direct orbit around a black hole. (In this work we always consider the direct orbits.)

The increase of gravity in the close vicinity of the rotating black hole in 
consequence leads to a smaller disk height, which will be given by the hydrostatic 
equilibrium:
\begin{equation}
H = \sqrt{P \over \rho} {1 \over \Omega} \sqrt {B \over C}.
\end{equation}

The angular velocity in the disk around a spinning black hole is given by:
\begin{equation}
\Omega = {c^{3} \over GM \left(\left({r c^{2} \over G M}\right)^{3/2}+a\right)}
\label{eq:omegad}
\end{equation}

The viscous shear will be modified as:
\begin{equation}
\tau_{r\varphi}=-\alpha P {A \over \sqrt{BC}}.
\end{equation}
Consequently, the heating and cooling terms in the energy balance equation will be
affected by the above correction factors.
Also, the equation for the energy flux generated in the disk, 
Eq. (\ref{eq:ftot}), will now be specified in the Kerr metric. Therefore, 
instead of the Newtonian boundary condition, we use the function:
\begin{equation}
f(r) = {D \over B}.
\label{eq:bdrycond}
\end{equation}
 This factor is equal to zero at the $r_{\rm ms}$ radius and approaches unity at large radii. This asymptotic behaviour of f(r) is the same as for the boundary condition 
derived in Chen \& Beloborodov (2007), who used more complex formalism in the Kerr metric. The torque imposed on the disk due to
 magnetic coupling with the black hole (see below)
does not change the no-torque inner boundary condition either.

\subsection{Magnetic heating}
\label{sec:qmc}

The spinning black hole may transfer its rotation energy to the jet via the 
Blandford-Znajek (BZ) process (Blandford \& Znajek 1977), as well as to 
the accretion disk, via the magnetic coupling (Li \& Paczy\'nski 2000). 
In the latter, the closed lines of magnetic field connect the black hole horizon
with the disk, and if the BH rotates faster than the surrounding disk, the
torque is exerted and the energy and angular momentum are extracted from the BH.
On the other hand, the direction of energy transfer may be reversed, if the disk 
angular velocity is larger than that of the black hole.

The details of the process depend on the structure of magnetic field. Here we follow the
assumptions of Wang et al. (2003), that the magnetic field is constant at the horizon
 and varies as a power law with the disk radius. We also use the mapping relation
between the angular coordinate, $\theta$, on the BH horizon and the disk radius, $r$,
derived by these authors. Due to this relation, the closed magnetic filed lines are 
attached not only to the inner edge of the disk, but connect the horizon region of 
smaller $\theta$ with outer disk radii, up to some maximum radius $r_{\rm out}$.
Therefore, the disk inner ring, between $r_{\rm ms}$ and $r_{\rm out}$, will
be affected by the energy transfer to and from the rotating black hole, 
depending on its spin and magnetic field configuration.
Because the inner regions of the disk are also subject to the thermal instability,
we intend to check what effect on this instability may arise due to the energy transfer 
between the black hole and the disk.

The mapping relation is defined by the following change of variables 
(Wang et al. 2003):
\begin{equation}
\sin \theta d\theta = - G(a,\xi, n) d\xi
\end{equation}
and
\begin{equation}
\cos \theta = \int_{1}^{\xi} G(a, \xi, n) d\xi
\end{equation}
where
\begin{equation}
G(a,\xi,n) = - { \xi^{1-n}\chi^{2}_{\rm ms}\sqrt{1+a^{2}\chi^{-4}_{\rm ms}\xi^{-2} + 2 a^{2}\chi^{-6}_{\rm ms}\xi^{-3}} \over 2 \sqrt{(1+a^{2}\chi^{-4}_{\rm ms}+2 a^{2}\chi^{-6}_{\rm ms})(1 - 2 \chi^{-2}_{\rm ms}\xi^{-1}+ a^{2}\chi^{-4}_{\rm ms}\xi^{-2})}}.
\end{equation}
Here the dimensionless radius is $\xi = r/r_{\rm ms}$, $\chi = \sqrt{r/r_{\rm g}}$ and
$r_{\rm g}=GM/c^{2}$. The mapping function $G(a, \xi, n)$ depends on the black hole 
spin, which is a parameter of our model, and on the strength of magnetic field at a given radius. For the magnetic field,  is assumed that it varies as (Blandford 1977):
\begin{equation}
B_{\rm z} \propto \xi^{-n}
\label{eq:bz}
\end{equation}
and the index $n$ is a model parameter.

The magnetic coupling between the rotating black hole and the disk
exerts a torque, equals to:
\begin{equation}
T_{\rm MC} = T_{0} 4a\left(1+\sqrt{1-a^{2}}\right) \int_{0}^{\pi/2} {(1-\beta)\sin^{3}\theta d\theta \over 2 - \left(1 - \sqrt{1-a^{2}}\right)\sin^{2}\theta },
\label{eq:Tmc}
\end{equation}
where $\beta = \Omega /\Omega_{H}$ is the ratio between the angular velocity in the disk, given by Eq. \ref{eq:omegad}, and the velocity at the black hole horizon, given by
\begin{equation}
\Omega_{\rm H} = a/\left(2r_{\rm g}\left(1+\sqrt{1-a^{2}}\right)\right).
\end{equation}
The integration over coordinate $\theta$ 
from $\theta_{\rm min}$ to $\theta_{\rm max}$
is equivalent to the integration over disk radius from $\xi_{\rm max}$ to $\xi_{\rm min}$. Here we set $\xi_{\rm min}=\xi_{\rm ms}=1$ at  $\theta = \pi/2$ and for simplicity 
we assume 
that no open magnetic field lines take the black hole rotation power to infinity, i.e.
the BZ effect is vanishing and 
$\xi_{\rm max}$ has its largest value for $\theta_{\rm min}=0$.

The normalization of the torque is (Wang et al. 2003):
\begin{equation}
T_{0} = 3.26\times 10^{45} \left({B_{\rm H} \over 10^{15} G}\right)^{2} \left({M \over M_{\odot}}\right)^{3} ~~~ {\rm g}~{\rm cm}^{2}~{\rm s}^{-2}
\end{equation}
and the energy density of the magnetic field 
is proportional to the total pressure in the disk (Moderski, Sikora \& Lasota 1998):
\begin{equation}
B^{2}_{\rm H} \propto 8\pi P^{\rm max} .
\end{equation}
In general, the proportionality coefficient in the above equation 
can be on the order of 1.0, however, some departures from the equipartition
of magnetic energy are possible. In case of the $\alpha$-disks, the ratio
of 
${B^{2}_{\rm H} / 8\pi P^{\rm max}} = \beta_{\rm mag}$
cannot be too high,
and will rather be on the order of $\alpha$ (Moderski, private communication).

Finally, the energy transfer from the rotating black hole to the accretion disk
results in the additional heating term:
\begin{equation}
Q^{+}_{\rm MC} = {T_{\rm MC} \Omega \over 4\pi r^{2}}.
\end{equation}

\section{Results}
\label{sec:results}

\subsection{Disk instability for a rotating black hole}
\label{sec:kerr_results}

We found that the thermal instability of the neutrino cooled disk, previously
studied for the case of Schwarzschild black holes in Janiuk et al. (2007), is
also present in the disks of Kerr black holes.
The instability arises in the inner parts of the disk, where the helium is 
fully photodisintegrated, the nuclei pressure is locally lower, while 
the neutrino opacity is large and the dominant pressure terms are due to the 
neutrinos and radiation. In our previous work, 
we found for a non-rotating black hole that the radial extension of the 
unstable strip was up to about 8 Schwarzschild radii. 
The accretion rate for which the instability
appeared, had to be extremely large, on the order of 12 solar masses per second.
The instability was manifested in both stationary models, as a phase 
transition to a distinct branch of distribution for various quantities 
(temperature, density, electron fraction, chemical potentials of species, etc.), 
as well as in the time dependent evolution of the disk (variability in time).

\begin{figure}
   \centering
\includegraphics[width=7cm]{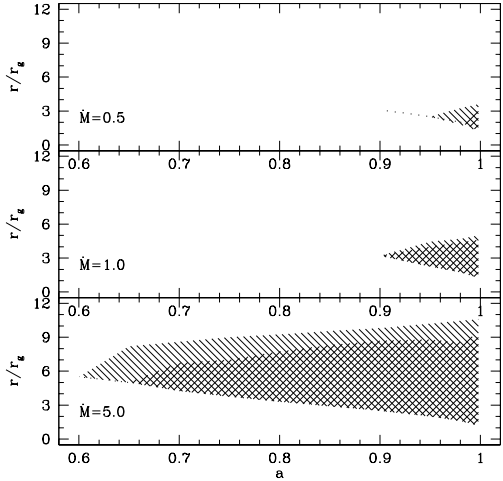}
\caption{The parameter space for which the unstable solution is found, 
in the plane radius - spin parameter. The bottom panel shows the results for 
accretion rate $\dot M=5.0$ $M_{\odot}$ s$^{-1}$, the middle panel is for 
$\dot M=1.0$ $M_{\odot}$ s$^{-1}$ and the top panel is for 
$\dot M=0.5$ $M_{\odot}$ s$^{-1}$. The
two values of viscosity were used: $\alpha=0.1$ (shaded area) and  $\alpha=0.3$ 
(cross-shaded area).
 } 
\label{fig:global}
\end{figure}

In this work, we calculated a grid of stationary models in the Kerr metric, 
to check for the 
location of the unstable strip depending on the black hole spin.
We used moderate accretion rates and several viscosity parameters.
In Figure \ref{fig:global} we show the radial extension of the
unstable strip, as a function of spin parameter $a$, for three values 
of accretion rate: $\dot M=0.5, 1.0$ and 5.0 $M_{\odot}$ s$^{-1}$. We find that 
for the extreme Kerr black hole,
the instability appears already at very low accretion rates, 
0.5 solar masses per second. 
The dependence on $\alpha$ is rather weak. In general, the larger is viscosity,
the larger BH spin is needed for the thermal instability to appear 
at a given accretion rate. Also, 
for larger alpha, the instability strip is narrower.
As shown in the upper panel of the Figure,
only for $a>0.95$ at viscosity $\alpha=0.1$, and for $a>0.98$ at $\alpha=0.3$
the unstable solutions are found at this low accretion rate.
For lower accretion rates, we checked  $\dot M=0.1$ $\dot M_{\odot}$ s$^{-1}$,
and no unstable solutions were found down to $\alpha=0.03$ and at $a=0.998$.

In general, the predicted trend is conserved: 
the higher the accretion rate and the larger black hole spin, the larger is 
the extension of the unstable strip. The trend with viscosity is the following: 
for larger viscosity parameters the extension 
of the unstable zone shrinks.

\begin{figure}
   \centering
\includegraphics[width=7cm]{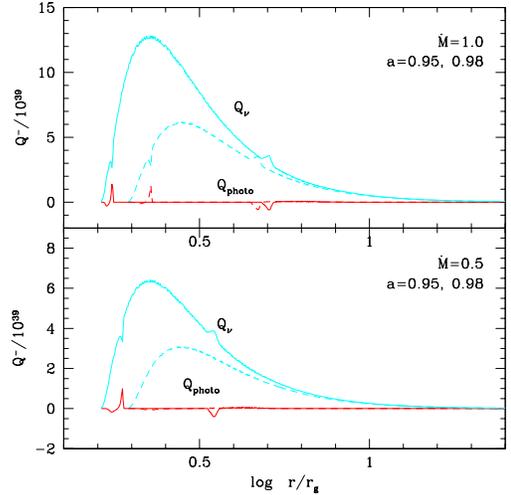}
\caption{The main cooling terms in the disk as a function of radius:
neutrino cooling (thick blue lines) and helium photodisintegration (thin red lines).
The bottom panel shows the results for 
accretion rate $\dot M=0.5$ $M_{\odot}$ s$^{-1}$, and the upper panel is 
for $\dot M=1.0$ $M_{\odot}$ s$^{-1}$. The
two values of black hole spin parameter are shown: $a=0.98$ (solid lines) and
$a=0.95$ (dashed lines). The viscosity parameter is $\alpha=0.1$.
 } 
\label{fig:cool}
\end{figure}

In the Figure \ref{fig:cool} we show the dominant cooling terms as a function of radius.
The terms are due to the neutrino emission, $Q_{\nu}$, and helium photodisintegration, 
$Q_{\rm photo}$. The other cooling terms, due to radiation and advection of energy, 
are incorporated in our model, but they are a few orders of magnitude smaller 
than neutrino cooling. Therefore we do not show these terms on the plot. 
The  $Q_{\rm photo}$ term at large radii 
can locally have a negative value, which means 
that the helium nuclei are at this radius created.

The unstable strip in the accretion disk can be localized between the two spikes in the
$Q_{\rm photo}$ distribution. The first spike is very close to the inner 
edge of the accretion disk, with a small offset because of 
the inner boundary condition (cf. Eq. \ref{eq:bdrycond}). Due to this condition, no torque is induced on 
the inner edge.
 This results in zero viscous dissipation rate and 
the plasma temperature drops close to the inner edge 
(in practice, for numerical reasons we set our inner radius of the computational domain
slightly above inner edge, to avoid singularities). Note, that the torque 
induced by the magnetic coupling between the disk and the black hole is transported 
only outwards with radius, while at the inner edge of the disk the torque is still zero.We also note, that the density drop at the inner edge is compensated by a large 
radial velocity, because we take into account the advective cooling. Therefore, despite the small disk thickness, the accretion rate is kept constant.
Above the inner edge of the torus, the temperature reaches 
$10^{10}-10^{11}$ K, while the density exceeds $10^{9}$ g cm$^{-3}$, 
and the helium nuclei 
can appear. The plasma consists of free neutrons, protons and electron-positron pairs,
and if the photon flux is sufficient, the helium nuclei are photodisintegrated.
Depending on the model parameters, the outer radius of this strip can 
reach up to 12 gravitational radii (Fig. \ref{fig:global}).
Outwards of this radius, the helium is again synthesized and its 
photodisintegration contributes to the energy balance in the outer disk (with some
small fluctuations in $Q_{\rm photo}$ up to $r_{\rm out}$, not visible in the 
Figure).

The neutrino cooling term is always a dominant one in all the models.
The larger accretion rate and black hole spin, the hotter is the disk and hence the
neutrino cooling is larger. For the rotating black hole, the inner disk radius shifts
closer (cf. Eq. \ref{eq:rms}), and also the maximum in the 
heating and cooling distribution is shifted inwards.

\begin{figure}
   \centering
\includegraphics[width=7cm]{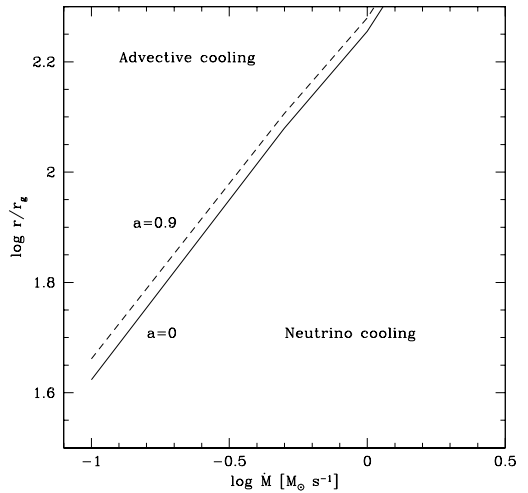}
\caption{The parameter space for which the disk is dominated by the neutrino cooling 
(NDAF).
We plot the maximum radius of NDAF (in the units of $r_{g}=GM/c^{2}$) 
as a function of accretion rate, for 2 
chosen values of black hole spin: a=0 (solid line) and a=0.9 (dashed line).
 } 
\label{fig:paramspace}
\end{figure}

We note, that our model works well in the neutrino-cooling dominated region, which
is always inside the radius $r_{\rm max}$. This 'neutrino ignition radius' 
strongly depends on the accretion rate
and weakly depends on the black hole spin. Outside this region, the flow is dominated by advection and neutrino emission is inefficient in cooling the disk (see also Chen \& Beloborodov 2007;  note however that we do not consider the gravitational instability which according to these authors is present at large radii and large accretion rates,
 nor we study the case of the smallest accretion rates, for which the neutrino cooling is inefficient throughout the disk at all radii).
In Figure \ref{fig:paramspace} we indicate the parameter space for 
which the neutrino cooling is dominant.
The instability we discuss here, as well as the magnetically coupled region 
we discuss below, are both located well inside the neutrino cooled region.

\subsection{Magnetic coupling between the disk and rotating black hole}
\label{sec:magnetic}

The mapping relation between the polar coordinate on the black hole 
horizon and the disk radius implies that there exists a maximum radius, defining the
region of influence of the rotating black hole. Below this radius, at 
$\xi<\xi_{\rm out}$,
the energy may be transferred to and/or from the black hole to the disk.
Above this radius, there are no closed magnetic field lines, and the disk is 
no longer subject to the heating by the black hole energy extraction.

\begin{figure}
   \centering
\includegraphics[width=7cm]{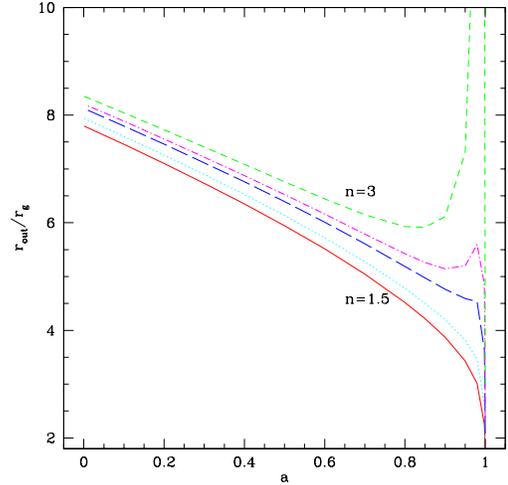}
\caption{The radial extension of the region threaded by magnetic coupling
between the disk and rotating black hole, as a function of its spin.
The solid line marks the power law index of magnetic field $n=1.5$, the dotted line is for $n=2$, long-dashed line is for $n=2.5$, the ditted-dashed line is for $n=2.7$
 and the short dashed line is for $n=3$.
 } 
\label{fig:xiout}
\end{figure}

In the Figure \ref{fig:xiout} we show the dependence of the radial extension of the 
magnetically coupled region on the black hole spin and magnetic field distribution.
The latter is parameterized by the index $n$ in Eq. \ref{eq:bz}, and the Figure shows 
several examples for $n$=1.5, 2.0, 3.0 and 6.0.
As was found by Wang et al. (2003), the
influence of the black hole rotation is concentrated in the closest 
vicinity of the inner disk edge and the radius $\xi_{\rm out}=r_{\rm out}/r_{\rm ms}$ 
is very small 
for small values of index $n$. For $n \le 2.0$  the largest $\xi_{\rm out}$
is about 2.2, at the black hole spin $a=0.98$ and then drops. 
The result for larger $n$ are quantitatively the same as in
 Wang et al. (2003) and they differ only quantitatively. For
$n=3$ the maximum is
$\xi_{\rm out} > 10.0$ at $a=0.998$, while those authors had a value about 
$\xi_{\rm out} \approx 4.0$, which in turn we obtained for $n=2.7$. 
The reason is probably the sensitivity of the results to the
 numerical method used to compute the integrals in Eq. \ref{eq:Tmc}.
For $n>3.0$ and large $a$, the size of the region $\xi_{\rm out}$ goes to infinity, and 
the larger $n$, the smaller is the critical spin parameter at which 
that happens (see Wang et al. 2002).

Note that actually in Fig. \ref{fig:xiout} we plot the results in the units of the 
gravitational radius, $r_{\rm g}$, instead of $r_{\rm ms}$ radius, 
i.e. the dependence on the black hole spin is accounted for also in the 
$r_{\rm ms}$ value. Therefore we can directly compare the extension of the 
magnetically coupled region with the extension of the thermal instability in 
the disk, presented in Fig. \ref{fig:global}.
In these units, the size of the magnetically coupled region decreases with $a$, 
because also the $r_{\rm ms}$ radius is smaller (inner edge is 
closer to the black hole) if the spin is large. 
For the most interesting cases of the large spin, the outer radius of magnetically 
coupled region is usually well inside the thermal instability strip, provided the accretion rate is sufficient for the thermal instability to occur, and $n$ is rather small.
However if $n \ge 3$, the magnetically coupled region may extend to larger radii than
the thermally unstable strip, especially for moderately small accretion rates.

\begin{figure}
   \centering
\includegraphics[width=7cm]{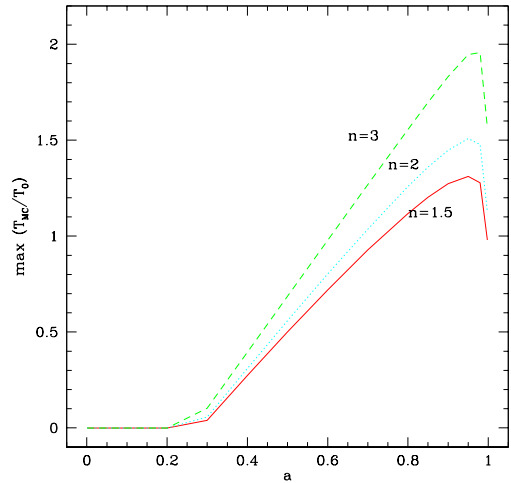}
\caption{The maximum magnetic torque induced in the disk, 
as a function of the black hole spin. The solid line marks the power law index 
of magnetic field $n=1.5$, the dotted line is for $n=2$, the long dashed line is
for $n=2.5$, the dotted-dashed line is for $n=2.7$
 and the short dashed line is for $n=3$.} 
\label{fig:maxtmc}
\end{figure}

However the size of the magnetically threaded strip may be small, the energy input
given to the disk by a rotating black hole is quite large.
In the Figure \ref{fig:maxtmc} we show the maximum magnetic torque 
induced in the disk by the black hole rotation, depending on its spin, for several values of index $n$.
The value of the torque increases with radius and 
saturates at $\xi_{\rm out}$ at its maximum value. This value is correlated 
to both index $n$ and spin, decreasing only for the extreme Kerr black holes.
The value of maximum $T_{\rm MC}=0$, obtained for the most slowly rotating black holes,
 means that in fact for these parameters the torque is negative throughout the 
magnetically coupled region. This means that
the rotation energy is transferred only from the disk to the black hole.
On the other hand, for the fast rotating black holes, and large $n$, 
the value of $T_{\rm MC}$ may be large and positive
at the outer parts of the magnetically coupled region, while
negative in the innermost radii. This is because if $r_{\rm ms}$ is sufficiently small, the innermost disk parts rotate faster 
than the black hole, 
and the energy is extracted from the inner disk to spin up the black hole.

Now, we study how the magnetic coupling between the disk and the rotating black hole
will influence the thermal-viscous instability, found by Janiuk et al. (2007).
This thermal instability was studied 
in case of the non-rotating black hole and found to require 
very high accretion rates. In the present 
work we confirmed the instability occurs also for the rotating black holes and
needs smaller accretion rates (Section \ref{sec:kerr_results}).
To further investigate the role of the black hole rotation, 
we calculated a grid of models parameterized by the accretion rate $\dot M$, 
black hole spin $a$, as well as the magnetic field power law index $n$ and
its normalization $\beta_{\rm mag}$, which describe the strength of 
the magnetic coupling.

\begin{figure}
  \centering
\includegraphics[width=7cm]{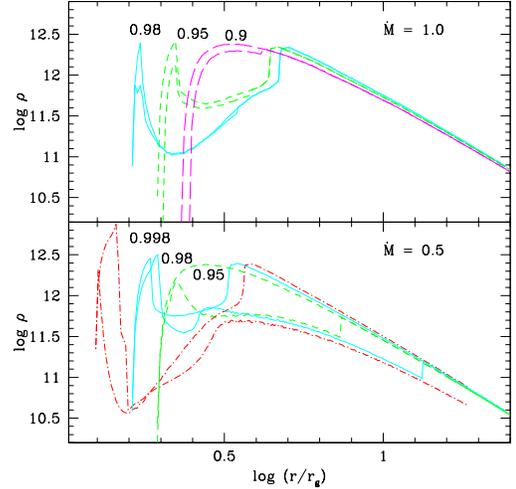}
\caption{The density profiles for the magnetically heated/cooled disks
(thick lines), for several values of the black hole spin:
$a=0.998$ (dot-dashed lines), $a=0.98$ (solid lines), $a=0.95$ (short dashed lines)
and $a=0.9$ (long dashed lines). The thinner lines show the case when the magnetic 
torque was neglected. The parameters were $\dot M = 1.0 M_{\odot}$s$^{-1}$ (top panel)
and  $\dot M = 0.5 M_{\odot}$s$^{-1}$ (bottom panel). The magnetic torque parameters
were $n=2$, $\beta_{\rm mag}=0.1$ (top panel), and  $n=3$, $\beta_{\rm mag}=0.5$ 
(bottom panel). The viscosity parameter is $\alpha=0.1$. 
} 
\label{fig:roqmc}
\end{figure}

In Figure \ref{fig:roqmc} we show the density profiles for two values of
accretion rate,  $\dot M = 0.5$ and 1.0 $M_{\odot}$s$^{-1}$, and several
spin parameters in the range from $a=0.9$ to 0.998. We concentrate on the innermost 
region of the accretion disk, where both the thermal-viscous instability
and the magnetic coupling overlap. Therefore the Figures show the region up to 
the radius $r= 15 r_{\rm ms}$.
As discussed in Janiuk et al (2007), the thermal-viscous instability results
in a local density decrease accompanied by the temperature rise in the inner 
strip of the disk. Outwards of that strip, a stable solution is found, for which
the density and temperature smoothly decrease with radius.

Now, we compare our previous unstable solutions, calculated under the assumption of
no magnetic torque 
(i.e. $\beta_{\rm mag}=0$, shown by the thinner lines), with the new solutions 
with the magnetic torque included (i.e. $\beta_{\rm mag}=0.5$ or 0.1, 
shown by the thicker lines). We note that the density dip is deeper, whenever
the energy is transferred from the black hole to the disk. Only very close 
to the inner edge of the disk, for large BH spins and index $n\sim 3$, we have the 
black hole extracting energy from the disk. When this is coupled to the thermal 
instability,
the density at the innermost radii increases, and the instability strip effectively 
shifts towards larger radii.
For the outer edge of the instability strip, we can identify it with the local 
maximum of the density. Therefore, this point is shifted towards a smaller radius and
in turn the whole unstable strip is narrower. However, if both $a$ and $n$ are 
sufficiently large, the magnetically coupled region extends well outside the
instability strip (cf. Fig.  \ref{fig:roqmc}; bottom panel). In this case,
the density profile is still affected, and the density is systematically lower than
in the respective model without the magnetic torque. However, now the density
 also decreases smoothly with radius, and we do not find
 this region unstable.

For smaller index, $n=2$, the magnetically coupled region does not 
cover the whole instability strip. This is shown in the top panel of the Figure 
\ref{fig:roqmc}. Therefore we have the density decrease only in some part of 
that strip, while outwards the solution matches the $Q_{\rm MC}-=0$ case.
The effect is in these models smaller than in case of $n=3$, because we now use
the $\beta_{\rm mag}=0.1$ instead of  $\beta_{\rm mag}=0.5$, so that the
magnetic heating does not exceed the total emitted flux $F_{\rm tot}$.
For $n=2$ and $a<0.98$, the magnetic 
torque is everywhere positive, i.e. there is no region 
where the energy is extracted from the disk towards the black hole, and there is 
no density rise on the inner edge.

The local density dip is produced by the magnetic term 
also in these models, where there is no thermal instability.
This is shown in Fig. \ref{fig:roqmc} for $\dot M = 0.5$, $a=0.95$ (bottom panel) and
for  $\dot M = 1.0$, $a=0.9$ (upper panel). The radial extension, 
depth and localization of this dip are determined by the parameters
$n$, $a$ and $\beta_{\rm mag}$.

\begin{figure}
  \centering
\includegraphics[width=7cm]{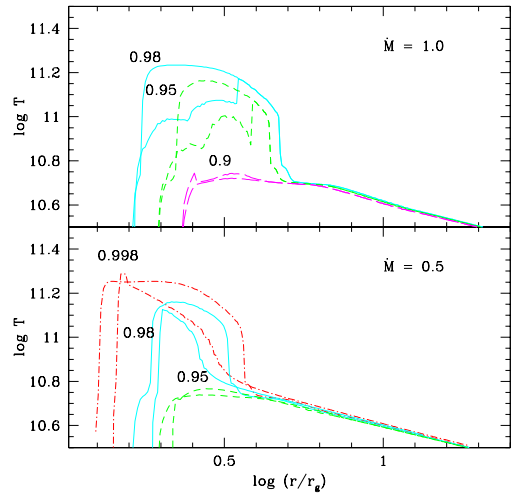}
\caption{The temperature profiles for the magnetically heated/cooled disks
(thick lines), for several values of the black hole spin:
$a=0.998$ (dot-dashed lines), $a=0.98$ (solid lines), $a=0.95$ (short dashed lines)
and $a=0.9$ (long dashed lines). The thinner lines show the case when the magnetic 
torque was neglected. The parameters were $\dot M = 1.0 M_{\odot}$s$^{-1}$ (top panel)
and  $\dot M = 0.5 M_{\odot}$s$^{-1}$ (bottom panel). The magnetic torque parameters
were $n=2$, $\beta_{\rm mag}=0.1$ (top panel), and  $n=3$, $\beta_{\rm mag}=0.5$ 
(bottom panel). The viscosity parameter is $\alpha=0.1$. 
} 
\label{fig:tqmc}
\end{figure}

In the Figure \ref{fig:tqmc} we plot the temperature profiles for 
the same models as were shown in Fig. \ref{fig:roqmc}.
As we mentioned above, the local density dip is accompanied by the temperature 
excess in the unstable region. However, when that region is magnetically 
threaded by a rotating black hole, the excess is smaller and extends to a 
smaller radius than for $Q_{\rm MC}=0$. In contrast, the stable disk 
outside the instability strip, is heated up and the temperature there  
increases, because the extra heating term is added.
On the other hand, the temperature decreases
close to the inner edge
in the models with a negative torque, i.e. when the energy is transferred 
from the disk to the black hole.
The stable disk affected by the magnetic coupling is therefore cooled at the radii where
$Q_{\rm MC}<0$ and heated at the region of $Q_{\rm MC}>0$.

Such a behaviour of the temperature profile, which shows the net 
decrease of the temperature when the magnetic heating is added in the unstable region, 
is caused by the nature of the thermal/viscous instability. As discussed 
in detail in Janiuk et al. (2007), the instability is a result of a phase 
transition, which occurs under certain conditions in the nuclear matter.
As a result, the extra energy provided by the rotation of the black hole 
via the magnetic coupling is consumed by this phase transition, and cannot heat the
disk.

We conclude, that the magnetic coupling between the spinning black hole
and the disk does influence the thermal-viscous instability.
It can result in shifting the unstable region
towards larger inner radius and shrinking it in size. 
The strength of the thermal-viscous instability
will be modified under the influence of the magnetic coupling,
 because the density dip in the unstable region will be deeper. 
However,
the temperature excess in the unstable region 
is smaller, if the disk magnetically coupled to the black hole. 
To quantify the net effect and 
calculate the resulting luminosity fluctuations we
would need to make the time-dependent modeling, which is beyond the scope of the 
present paper. We intend to study this in more detail in the future work.

We also note that there is some small
impact of magnetic coupling on the stable disk solutions.
This is an extra dip in density accompanied with the temperature excess,
which arises in the previously stable region.
However this effect seems to be much less pronounced than the 
thermal-viscous instability. This may potentially lead to the development
of a separate type of a magnetic instability.

\section{Discussion}
\label{sec:diss}

 The importance of black hole spin in the accretion disk systems and
the formation of jets, also in the presence of the magnetic fields,
has been studied by many authors.
It has been discussed e.g. in the review by Blandford (1999).
In the standard accretion disk theory (Shakura \& Sunyaev 1973)
it is assumed that there is no coupling between the black hole and the disk.
However, if the magnetic field is considered, such a coupling may be present.
Contrary to the Blandford-Znajek process (in which the magnetic filed lines 
threading the black hole may close on a very remote load, far away from the black 
hole), the magnetic field lines coupled to the accretion disk are closed in the 
vicinity of the inner radius.
In this process, 
the energy and angular momentum are extracted from the black hole
and transferred to the disk, if the hole rotates faster than the disk 
(MacDonald \& Thorne 1982; Li 2000). Such coupling may have much higher
efficiency in extracting the black hole rotation energy than the 
Blandford-Znajek mechanism, as was first quantitatively estimated by Li (2002).

Let us also remind here after Li (2002), that 
in this model 
the torque produced by the magnetic coupling with the black hole propagates only 
outwards, and no torque is imposed at the inner boundary of the disk. 
The non-zero torque at the inner boundary 
is possible if the disk is magnetically connected with the plasma 
in the transition region between the disk and the black hole (e.g. Krolik 1999), 
which we do not consider here.
Another important assumption behind the present model 
is that about the quasi steady-state of the disk and 
magnetic field configuration. Under this assumption, the magnetic field lines are 
not tangled by the rotation and MHD instabilities of the Balbus \& Hawley (1991) 
are not taken onto account. In our present work we hold this assumption, after
Li (2002) and Wang et al. (2002; 2003a,b).
The evolution of the coronal magnetic field
threading the differentially rotating disk was studied however in the literature,
 e.g. by Lovelace et al. (2002)
in the context of Poynting flux dominated jets in quasars, and in Uzdensky (2004; 2005)
by numerically solving the Grad-Shafranov equation in the Schwarzschild and Kerr metric.

In the context of Gamma Ray Bursts, the black hole spin interaction with 
the magnetized torus  
was studied by van Putten et al. (2004)
These authors take
into account the dissipation of energy in the torus not only via the thermal and 
neutrino emission, but also through the gravitational waves and winds. They estimate 
the lifetime of the central engine powered by rapidly spinning black hole, 
whose spin-down is governed by the surrounding magnetic field coupled to the 
inner torus.
In the present model, we neglect the gravitational waves emission and we concentrate
on the structure of the neutrino-cooled disk. We restrict ourselves to the
steady-state solution, however we note that the lifetime of the
long gamma ray burst central engine is defined by the period of the rapid spin of the 
black hole, which may decrease with time during the engine activity (see also Janiuk et al. 2008).

In this article, we study the model of a stationary 
neutrino cooled disk around a rotating Kerr black hole. We confirmed the
presence of the thermal-viscous instability in the region of the 
large neutrino opacity and efficient helium photodisintegration. 
The instability was first studied in Janiuk et al. (2007) 
in the time-dependent model for the Schwarzschild black hole. 
It
was found to produce dramatic changes in the density and 
temperature profiles, as well as a rapid time variability of the disk emission.
Here, we quantitatively described the role of the black hole spin 
and viscosity in the disk. 
We found that in the Kerr black hole disks the large spin parameter results in
a larger size of an unstable strip, which appears even for a moderate accretion rate.
The critical accretion rate is anticorrelated with the BH spin required for the 
onset of the instability.
In the extreme BH case, the instability close to the inner edge may be effective 
even for $\dot M=0.5 M_{\odot}$ s$^{-1}$. On the other hand, 
the more viscous is the disk, the smaller in size is the unstable strip and the weaker
should be the instability. 

The time dependent calculations 
were not performed in the present work, because they would require to compute
 time evolution of
the BH spin and a moving radial grid, which is intended to be studied in the 
future work. However, we studied the influence of the black hole rotation on the disk 
structure considering our stationary model. This was done by means of
the magnetic transfer of energy from the rotating black hole to the disk, 
and vice versa.

The first important finding in this work is that 
thermal instability occurs for quite low accretion rates, on the order of a 
fraction of solar mass per second, provided the black hole rotates very fast.
This can be relevant for the origin of the variable energy input to the GRB jets
not only in case of the short GRBs, but also for long ones. The latter are 
presumably powered by the accretion of the fallback material from the massive star
 envelope onto the newly 
born black hole after the hypernova explosion (e.g. Mac Fadyen \& Woosley 1999).
In these models, the accretion rate is found to be substantially lower than in the 
merger of two neutron stars, and therefore the event can last for much longer time to 
power the GRB for several hundreds of seconds.
On the other hand, the massive star that is a progenitor of a 
hypernova is a rotating star
and therefore a natural consequence of the explosion should be the 
rapidly rotating black hole in the center.

As was recently shown by Janiuk, Moderski \& Proga (2008), 
the large black hole spin is a key ingredient for the occurrence of the 
longest duration GRBs, and is required for the GRB central engine to operate 
for a time of the order of hundreds of seconds.
The present study leads us therefore to a general conclusion that the
longest GRBs, that are powered first by the neutrino annihilation
and later by the black hole spin and require $a>0.9$, should be very variable
at the initial phase of the prompt emission. Later on, when the accretion rate drops, 
the variability may be suppressed and what is observed is a smooth 
tail in the gamma ray lightcurve. On the other hand, 
the period of the highly variable GRB emission may be somewhat extended 
if the black hole is being spun up by the accretion. 
The issue discussed in the literature was whether in such a case the spin 
equilibrium value can be achieved at $a=0.998$ (Thorne 1974), 
or a less value (Gammie, Shapiro \& McKinney 2004).

In our study, we also considered the heating of the innermost regions of the 
torus due to the magnetic coupling with the rotating black hole.
We found
 that this process can have some influence on the thermal instability, 
because the size and position of the unstable region are somewhat changed. Also, 
the profiles of density and temperature in the unstable region are modified.
The magnetic coupling enhances the change in density profile, however it 
weakens the change in temperature profile. The energy transfer to the disk
 from the rotating black hole does not suppress the thermal instability.
This finding also supports our conclusion that the variable and long duration 
GRBs are powered by the rapidly spinning black holes.

\begin{acknowledgements}
This work was partially supported by the
Polish Astroparticle Network grant 621/E-78/SN-0068/2007.
 This work is also partially supported by
 National Basic Research Program of China (grant 2009CB824800), the
 National Natural Science Foundation (grant 10733010,10673010,10573016).
\end{acknowledgements}

\end{document}